\documentclass{article}

\usepackage{graphicx}

\def\be{\begin{equation}}
\def\ee{\end{equation}}
\def\bea{\begin{eqnarray}}
\def\eea{\end{eqnarray}}

\newcommand{\e}{{\rm e}}

\newcommand{\lb}{\label}
\renewcommand{\d}{{\rm d}}

\title{Response Variability in Balanced Cortical Networks}
\author{
Alexander Lerchner \\
Technical University of Denmark, 2800 Lyngby, Denmark
\and
Cristina Ursta \\
Niels Bohr Institut, Blegdamsvej 17, 2100 Copenhagen {\O}, Denmark
\and
John Hertz \\
Nordita, Blegdamsvej 17, 2100 Copenhagen {\O}, Denmark \and
Mandana Ahmadi \\
Nordita, Blegdamsvej 17, 2100 Copenhagen {\O}, Denmark
\and Pauline Ruffiot \\
Universit\'e Joseph Fourier, Grenoble, France}

\begin{document}

\maketitle

\begin{abstract}
We study the spike statistics of neurons in a network with
dynamically balanced excitation and inhibition. Our model,
intended to represent a generic cortical column, comprises
randomly connected excitatory and inhibitory leaky
integrate--and--fire neurons, driven by excitatory input from an
external population. The high connectivity permits a mean-field
description in which synaptic currents can be treated as Gaussian
noise, the mean and autocorrelation function of which are
calculated self-consistently from the firing statistics of single
model neurons.  Within this description, we find that the
irregularity of spike trains is controlled mainly by the strength
of the synapses relative to the difference between the firing
threshold and the post-firing reset level of the membrane
potential.  For moderately strong synapses we find spike
statistics very similar to those observed in primary visual
cortex.
\end{abstract}

\section{Introduction}

The observed irregularity and relatively low rates of the firing
of neocortical neurons suggest strongly that excitatory and
inhibitory input are nearly balanced.  Such a balance, in turn,
finds an attractive explanation in the mean-field descriptions of
Amit and Brunel \cite{AmitBrunelCC,AmitBrunelNetwork,BrunelJCNS}
and Van Vreeswijk and Sompolinsky \cite{vVSScience, vVSNC}. In
their theories, the balance does not have to be put in ``by
hand''; rather, it emerges self-consistently from the network
dynamics. This success encourages us to study firing correlations
and irregularity in models like theirs in greater detail. In
particular, we would like to quantify the irregularity and
identify the parameters of the network that control it.  This is
important because one can not extract the signal in neuronal spike
trains correctly without a good characterization of the noise.
Indeed, an incorrect noise model can lead to spurious conclusions
about the nature of the signal, as demonstrated by Oram {\em et
al} \cite{Orametal}.

Response variability has been studied for a long time in primary
visual cortex
\cite{HeggelundAlbus,Dean,TMT,TMD,Vogelsetal,Snowdenetal,Guretal,ShadlenNewsome,Gershonetal,Karaetal,
Buracasetal} and elsewhere
\cite{Leeetal,Gershonetal,Karaetal,DeWeeseetal}. Most, though not
all, of these studies found rather strong irregularity.  As an
example, we consider the findings of Gershon {\em et al}
\cite{Gershonetal}.  In their experiments, monkeys were presented
with flashed, stationary visual patterns for several hundred ms.
Repeated presentations of a given stimulus evoked varying numbers
of spikes in different trials, though the mean number (as well as
the PSTH) varied systematically from stimulus to stimulus.  The
statistical objects of interest to us here are the distributions
of single-trial spike counts, for given fixed stimuli.  Often one
compares the data with a Poisson model of the spike trains, for
which the count distribution $P(n) = m^n \e^{-m}/n!$. This
distribution has the property that its mean $\langle n \rangle =
m$ is equal to its variance $\langle \delta n^2 \rangle = \langle
(n- \langle n \rangle )^2 \rangle$. However, the experimental
finding was that the measured distributions were quite generally
wider than this: $\langle \delta n^2 \rangle
> m$.  Furthermore, collecting data for many stimuli, the variance
of the spike count was fit well by a power law function of the
mean count: $\langle \delta n^2 \rangle \propto m^y$, with $y$
typically in the range $1.2 - 1.4$, broadly consistent with the
results of many of the other studies cited above.

Some of this observed variance could have a simple explanation:
The condition of the animal might have changed between trials, so
the intrinsic rate at which the neuron fires might differ from
trial to trial, as suggested by Tolhurst {\em et al} \cite{TMT}.
But it is far from clear whether all the variance can be accounted
for in this way.  Moreover, there is no special reason to take a
Poisson process as the null hypothesis, so we don't even really
know how much variance we are trying to explain.

In this paper, we try to address the question of how much
variability, or more generally, what firing correlations can be
expected as consequence of the intrinsic dynamics of cortical
neuronal networks.  The theories of Amit and Brunel and of van
Vreeswijk and Sompolinsky do not permit a consistent study of
firing correlations. The Amit-Brunel treatment assumes that the
input to neurons is uncorrelated in time (white noise).  Thus,
although one can calculate the variability of the firing
\cite{BrunelJCNS}, it is not self-consistent.  Van Vreeswijk and
Sompolinsky use a binary-neuron model with stochastic dynamics
which makes it difficult, if not impossible, to study temporal
correlations that might occur in networks of spiking neurons.
Therefore, in this paper we do a complete mean-field theory for a
network of leaky integrate-and-fire neurons, including, as
self-consistently-determined order parameters, both firing rates
and autocorrelation functions.  A general formalism for doing this
was introduced by Fulvi Mari \cite{FulviMari} and used for an
all-excitatory network; here we employ it for a network with both
excitatory and inhibitory neurons. A preliminary study of this
approach for an all-inhibitory network was presented previously
\cite{Hertz+Richmond+Nilsen:2003}.

\section{Model and Methods}

\begin{figure}[t]
\centering
 \includegraphics[width=9cm]{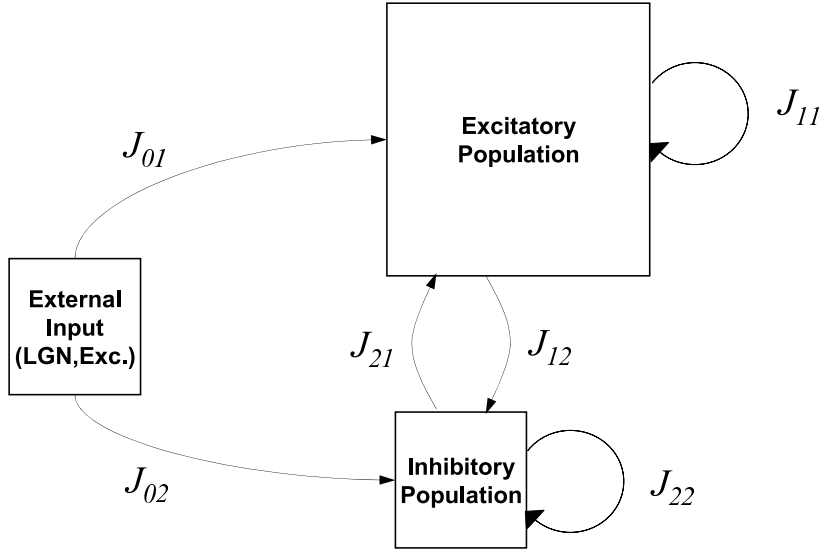}
 \caption{Structure of the model network.} \label{fig:Network}
\end{figure}

The model network, indicated schematically in
Fig.~\ref{fig:Network}, consists of $N_1$ excitatory neurons and
$N_2$ inhibitory ones.  In this work we use leaky
integrate-and-fire neurons, though the methods could be carried
over directly to networks of other kinds of model neurons, such as
conductance-based ones. They are randomly interconnected by
synapses, both within and between populations, with the mean
number of connections from population $b$ to population $a$ equal
to $K_b$, independent of $a$.  In specific calculations, we have
used $K_1$ from 400 to 6400, and we take $K_2 = K_1/4$. The
population sizes $N_a$ do not enter directly in the mean field
theory, only their ratios (the connection probabilities)
$K_a/N_a$. We have used $K_a/N_a = 0.1$ for both excitatory and
inhibitory connections, implying $N_1 = 4N_2$.

We scale the synaptic strengths in the way van Vreeswijk and
Sompolinsky did\cite{vVSScience,vVSNC}, with each nonzero synapse
from population $b$ to population $a$ having the value
$J_{ab}/\sqrt{K_b}$.  The parameters $J_{ab}$ are taken to be of
order 1, so the net input current to a neuron from the $K_b$
neurons in population $b$ connected to it is of order
$\sqrt{K_b}$.  With this scaling, the fluctuations in this current
are of order 1.

Similarly, we assume that the external input to any neuron is the
sum of $K_0 \gg 1$ contributions from individual neurons (in the
LGN, if we are thinking about modeling V1), each of order
$1/\sqrt{K_0}$, so the net input is of order $\sqrt{K_0}$.  In our
calculations, we have used $K_0 = K_1$.

We point out that this scaling is just for convenience in thinking
about the problem.  In the balanced asynchronous firing state, the
large excitatory and inhibitory input currents nearly cancel,
leaving a net input current of order 1.  Thus, for this choice,
both the net mean current and its typical fluctuations are of
order 1, which is convenient for analysis. The physiologically
relevant assumptions are only that excitatory and inhibitory
inputs are separately much larger than their sum and that the
latter is of the same order as its fluctuations.

Our synapses are not modeled as conductances.  Our synaptic
strength simply defines the amplitude of the postsynaptic current
pulse produced by a single presynaptic spike.

The model is formally specified by the sub-threshold equations of
motion for the membrane potentials $u_i^a$ ($a = 1,2$, $i=1,
\ldots N_a$): \be \frac{\d u_i^a}{\d t} = -\frac{u_i^a}{\tau}
+\sum_{b=0}^2 \sum_{j=1}^{N_b} J_{ij}^{ab}
 S_j^b(t),                        \label{eq:model} \ee together with the
condition that when $u_i^a$ reaches the threshold $\theta_a$, the
neuron spikes and the membrane potential is reset to a value
$u_r^a$.   The indices $a$ or $b = 1$ or 2 label populations:
$b=0$ refers to the (excitatory) population providing the external
input, $a=1$ refers to the excitatory population and $a=2$ to the
inhibitory population. In (\ref{eq:model}), $\tau$ is the membrane
time constant (taken the same for all neurons, for convenience),
the strength of the synapse from neuron $j$ in population $b$ to
neuron $i$ in population $a$ is denoted by $J_{ij}^{ab}$, and
$S_i^b(t) = \sum_s \delta(t-t_{jb}^s)$ is the spike train of
neuron $j$ in population $b$. We have ignored transmission delays,
and we take the thresholds $\theta_a = 1$ and the reset levels
$u_r^a$ equal to the rest value of the membrane potential, 0. In
our calculations, the thresholds are given a Gaussian distribution
with a standard deviation equal to 10\% of the mean. Analogous
variability in other single-cell parameters (such as membrane time
constants) could also be included in the model, but for simplicity
we do not do so here.

We assume that the neurons in the external input population
($b=0$) fire as independent Poisson processes.  However, the
neurons in the network ($b=1,2$) are not in general Poissonian; it
is their correlations that we want to find in this investigation.

\subsection*{Mean Field Theory: Stationary States}

We describe the mean field theory and its computational
implementation first for the case of stationary rates. When the
connectivity is large and random, as we will assume here, each of
the three terms in the sum on $b$ on the right-hand side of
(\ref{eq:model}) can be treated as a Gaussian random function with
time-independent mean.  The simplest case is $b=0$, the external
input.  For simplicity, we assume that all $N_0$ neurons in the
external population fire at the same rate, $r_0$.  But because of
the random connectivity, the net time-averaged input current they
provide to a neuron in cortical population $a$ can vary from
neuron to neuron.  Assuming large, dilute connectivity ($K_0 \gg
1$ and $K_0 \ll N_0$), the central limit theorem then implies \be
\langle I_i^{a0}(t) \rangle = \sum_j J_{ij}^{a0} r_0 = \sum_j
(\overline{J_{ij}^{a0}} + \delta J_{ij}^{a0})r_0 =
J_{a0}r_0(\sqrt{K_0}+ x_i^{a0}), \lb{eq:invar} \ee where
$x_i^{a0}$ is a Gaussian-distributed random number of unit
variance.  By $\langle \cdots \rangle$ we mean a time average or,
equivalently, an average over ``trials" (independent repetitions
of the Poisson processes defining the input population neurons).
We will generally use a bar over a quantity to indicate an average
over the neuronal population or over the distribution of the
$J_{ij}^{ab}$. (Note that these two kinds of averages are very
different things.)

Writing the spike train $S_j^0(t)$ for neuron $j$ in the input
population as \be S_j^0(t) = r_0 + \delta S_j^0(t),
\lb{eq:rplusdS} \ee with $\langle \delta S_j^0(t) \rangle = 0$, we
can write the fluctuations around $\langle I_i^{a0} \rangle$ as
\be \delta I_i^{a0}(t) = \sum_j J_{ij}^{a0} \delta S_j^0(t) =
J_{a0} \xi_i^{a0}(t) \lb{eq:innoise} \ee where $\xi_i^{a0}(t)$ is
white noise of power $r_0$: \be \langle \xi_i^{a0}(t)
\xi_i^{a0}(t') \rangle = r_0 \delta(t-t') \lb{eq:innoisecorr} \ee
Thus, quite generally, the input has a large mean value, of order
$\sqrt{K_0}$, plus Gaussian fluctuations of order 1. The
fluctuations are of two kinds.  One is constant for a given
neuron, independent of time and trial and arises from the fact
that the connectivity is random and the neurons in the input
population have a distribution of rates.  The other fluctuation is
a dynamical one, with correlations (independent of $i$) reflecting
the Poisson dynamics of the input population neurons.

The recurrent input terms $I_i^{ab}(t)$ also have large means and
fluctuations, static and dynamic, of order 1, but certain features
of their statistics are slightly different, as a systematic formal
derivation \cite{FulviMari,KZ} proves.  Here we do not give the
derivation, but just describe the result, which is that
$I_i^{ab}(t)$ can be written \be I_i^{ab}(t) = J_{ab}[ \sqrt{K_b}
r_b + B_b x_i^{ab}+ \sqrt{1-K_b/N_b}\xi_i^{ab}(t)],
\label{eq:decrec} \ee with \be r_b = \overline{r_j^b} =
\frac{1}{N_b}\sum_j r_j^b. \lb{eq:meanrate} \ee the average rate
in population $b$, $x_i^{ab}$ a unit-variance Gaussian random
number, \be B_b = \sqrt{\left( 1-\frac{K_b}{N_b}\right)
\overline{( r_j^b)^2}} \lb{eq:Bb} \ee and \be \langle
\xi_i^{ab}(t) \xi_i^{ab}(t') \rangle = C_b(t-t'). \lb{eq:recnoise}
\ee Here $C_b(t-t')$ is the average autocorrelation function of
the firing of neurons in population $b$, \be C_b(t-t') =
\frac{1}{N_b}\sum_j \langle \delta S_j^b(t)\delta S_j^b(t')\rangle
, \lb{eq:corrfn} \ee with \be \delta S_j^b(t)= S_j^b(t)-r_j^b .
\lb{eq:dSjb} \ee Again, $x_i^{ab}$ is time- and trial-independent,
while the noise $\xi_i^{ab}(t)$ varies both in time within a trial
and randomly from trial to trial.  Note that for this model a
correct and complete mean field theory has to include rate
fluctuations, through $\overline{(r_j^b)^2}$, and the firing
correlations, given by $C_b(t-t')$, as well as the mean rates.

The means of the recurrent input currents $I_i^{ab}$ are
completely analogous to the mean term in $I_i^{a0}$, but the
effective noise is different in three ways \begin{enumerate} \item
The amplitude $B_b$ of the static noise component (the second
term) contains a factor of the rms rate $\sqrt{\overline{(
r_j^b)^2}}$, not $r_b$ as in (\ref{eq:invar}). The same would be
true for the static input noise ($b=0$) if we allowed a
distribution of rates in the input population.  So this difference
is not an essential one.  It occurs only because we made a
simplifying assumption about the input population.  However, we
are not allowed to assume that about the neurons in the cortical
network, which will always have a distribution of rates because of
the random connectivity. \item The neurons providing the source of
these currents are not generally Poissonian, so their correlations
appear in the statistics of the noise term. \item The noise terms,
both static and dynamic, have a factor $\sqrt{1-K_b/N_b}$ in front
of them. This can be understood in the following way:  It is the
randomness in the synaptic connections in the network that
generates these noise terms in the effective single-neuron
problem; in general, they are proportional to $\overline{(\delta
J_{ij}^{ab})^2}$, which is equal to $J_{ab}^2 (1-K_b/N_b)/N_b$ in
our model. In the limit of full connectivity, $K_b = N_b$, all
$J_{ij}^{ab}$ are equal and there is no randomness. Therefore
there is no noise, as guaranteed here by this
factor.\end{enumerate}

The self-consistency equations of mean field theory are simply the
conditions that the average output statistics of the neurons,
$r_a$, $\overline{(r_j^a)^2}$ and $C_a(t-t')$ are the same as
those used to generate the inputs for single neurons using
integrate-and-fire neurons with synaptic input currents given by
(\ref{eq:invar}), (\ref{eq:innoise}) and (\ref{eq:decrec}).

In an equivalent formulation, the second term in
(\ref{eq:decrec}) can be omitted if the noise terms
$\xi_i^{ab}(t)$ have correlations equal to the unsubtracted
correlation function \be C_b^{\rm tot}(t-t') =\frac{1}{N_b}\sum_j
\langle S_j^b(t) S_j^b(t')\rangle \lb{eq:Ctot} \ee instead of
(\ref{eq:corrfn}).   For $|t-t'| \rightarrow \infty$, $C_b^{\rm
tot}(t-t') \rightarrow \overline{(r_j^2)}$, so $\xi_i^{ab}(t)$
acquires a random static component of mean square value
$\overline{(r_j^b)^2}$.

In still another way to do it, one can use the square of the
average rate, $r_b^2$ in place of $\overline{( r_j^b)^2}$ in
Eq.(\ref{eq:Bb}) for $B_b$ and employ noise with correlation
function \be \tilde C_b(t-t') =\frac{1}{N_b}\sum_j \langle
(S_j^b(t)-r_b)(S_j^b(t')-r_b)\rangle. \lb{eq:Ctilde} \ee For
$|t-t'| \rightarrow \infty$, \be \tilde C_b(t-t')\rightarrow
\overline{(r_j^b -r_b)^2} \equiv \overline{(\delta r_j^b)^2}.
\lb{eq:Ctildelimit} \ee  There are now two static random parts of
$I_i^{ab}(t)$, one from the $B_b$ term and one from the static
component of the noise.  Their sum is a Gaussian random number
with standard deviation equal to $B_b$ as given in
(\ref{eq:decrec}).  Thus these three ways of generating the input
currents are all equivalent.

\subsubsection*{The balance condition}

In a stationary, low-rate state, the mean membrane potential
described by (\ref{eq:model}) has to be stationary.  If excitation
dominates, we have $\d u_i^a/\d t \propto \sqrt{K_0}$, implying a
firing rate of order $\sqrt{K_0}$ (or  one limited only by the
refractory period of the neuron).  If inhibition dominates, the
neuron will never fire.  The only way to have a stationary state
at a low rate (less than one spike per membrane time constant) is
to have the excitation and inhibition nearly cancel.  Then the
mean membrane potential can lie a little below threshold, and the
neuron can fire occasionally due to the input current
fluctuations.  Thus, using (\ref{eq:invar}) and (\ref{eq:decrec}),
we have \be \sum_{b=0}^2  J_{ab}\sqrt{K_b}r_b =  {\cal O}(1)
\lb{eq:balance} \ee or, up to corrections of ${\rm
O}(1/\sqrt{K_0})$, \be \sum_{b=0}^2 \hat J_{ab} r_b = 0
\lb{eq:normbal} \ee with $\hat J_{ab} = J_{ab} \sqrt{K_b/K_0}$.
These are two linear equations in the two unknowns $r_a$, $a =
1,2$, with the solution \be r_a = \sum_{b=1}^2 [{\sf \hat
J}^{-1}]_{ab} J_{b0}r_0 ,                  \lb{eq:balsoln} \ee
where ${\sf \hat J}^{-1}$ is the inverse of the $ 2 \times 2$
matrix with elements $\hat J_{ab}$, $a,b = 1,2$.  If there is a
stationary balanced state, the average rates of the excitatory and
inhibitory populations are given by (\ref{eq:balsoln}) (in the
large-$N$ limit).

This argument depends only on the rates, not on the correlations,
and is exactly the same as that given by Amit and Brunel and by
Sompolinsky and van Vreeswijk.

This calculation does not say whether this state is stable,
however.  To determine this, one can expand around this solution
and examine the linear stability of the fluctuations, as done for
their model by van Vreeswijk and Sompolinsky \cite{vVSNC}.  Here,
we do not do this analytically, but rather check the stability of
our states numerically within our algorithm.

\subsubsection*{Numerical procedure}

For integrate-and-fire neurons in a stationary state, the mean
field theory can be carried out analytically if a white-noise
(Poisson firing) approximation is made
\cite{AmitBrunelCC,AmitBrunelNetwork,BrunelJCNS}.  But if firing
correlations are to be taken into account, it is necessary to
resort to numerical methods.  Thus we simulate single neurons
driven by Gaussian synaptic currents, collect their firing
statistics to compute the rates $r_a$, rate fluctuations
$\overline{(\delta r_j^a)^2}$ and correlations $C_a(t-t')$, and
then use these to generate improved input current statistics.  The
cycle is repeated until the input and output statistics are
consistent.  This algorithm was first used by Eisfeller and Opper
\cite{EisfellerOpper} to calculate the remanent magnetization of a
mean field model for spin glasses.

Explicitly, we proceed as follows.  We simulate single excitatory
and inhibitory neurons over ``trials'' 100 integration timesteps
long.  (We will call each timestep a ``millisecond''.  We have
explored using smaller timesteps and verified that there are no
qualitative changes in the results.)  We start from estimates of
the rates given by the balance condition, which makes the net mean
input current vanish.  Then the sum of the ${\cal O} (\sqrt{K_b})$
terms in (\ref{eq:invar}) and (\ref{eq:decrec}) vanishes, leaving
only the rate fluctuation and noise terms.   We then run 10000
trials of single excitatory and inhibitory neurons, selecting on
each trial random values of $x_i^{ab}$ and $\xi_i^{ab}(t)$.  Since
at this point we do not have any estimates of either the rate
fluctuations $\overline{(\delta r_j^b)^2}$ or the correlations
$C_b(t-t')$, we use $r_b^2$ in place of $\overline{( r_j^b)^2}$ in
Eq.(\ref{eq:Bb})for $B_b$   and use white noise for
$\xi_i^{ab}(t)$: $C_b(t-t') \rightarrow r_b \delta(t-t')$.

The random choice of $x_i$ from trial to trial effectively samples
across the neuronal populations, so we can then collect the
statistics $r_a$, $\overline{(r_j^a)^2}$ (or, equivalently,
$\overline{(\delta r_j^a)^2}$), and $C_a(t-t')$ from these trials.
These can be used to generate an improved estimate of the input
noise statistics to be used in (\ref{eq:decrec}) in a second set
of trials, which yields new spike statistics again. This procedure
is iterated until the input and output statistics agree. This may
take up to several hundred iterations, depending on network
parameters and how the computation is organized.

If one tries this procedure in its naive form, i.e., using the
output statistics directly to generate the input noise at the next
step, it will lead to big oscillations and not converge.  It is
necessary to make small corrections (of relative order $1/K_0$) to
the previous input noise statistics to guarantee convergence.

When one computes statistics from the trials in any iteration, the
simplest procedure involves calculating not (\ref{eq:corrfn}), but
rather $\tilde C_b(t-t')$ (Eq.(\ref{eq:Ctilde})).  From it, we can
proceed in two ways. In the first, from its $|t-t'| \rightarrow
\infty$ limit we can obtain $\overline{(\delta r_j^b)^2}$, and
thereby $\overline{(r_j^b)^2} = r_b^2 + \overline{(\delta
r_j^b)^2}$ for use in calculating $B_b$ in (\ref{eq:Bb}).
Subtracting this limiting value from $\tilde C_b(t-t')$ give us
$C_b(t-t')$ (which vanishes for large $|t-t'|$) for use in
generating the noise $\xi_i^{ab}(t)$.  This is the first of the
three methods described above.

Alternatively, we can use the third method:  At each step of our
iterative procedure we can generate noise directly with the
correlations $\tilde C_b(t-t')$ (which are long-ranged in time)
and use $r_b^2$ in place of $\overline{(r_j^b)^2}$ in calculating
$B_b$ (\ref{eq:Bb}). We have verified that the two methods give
the same results when carried out numerically, though the second
procedure converges more slowly.

While the true rates in the stationary case are time-independent
and $C_a(t,t')$ is a function only of $t-t'$, the statistics
collected over a finite set of noise-driven trials will not
exactly have these stationarity properties.  Therefore we improve
the statistics and impose time-translational invariance by
averaging the measured $r_a(t)$ and $\overline{(\delta
r_j^a(t))^2}$ over $t$ and averaging over the measured values
$C_a(t,t')$ with a fixed $t-t'$.

After the iterative procedure converges, so that we have a good
estimate of the statistics of the input, we want to run many
trials {\em on a single neuron} and compute its firing statistics.
This means that {\em the numbers $x_i^{ab}$ ($b=0,1,2$) should be
held constant over these trials}.  In this case it is necessary to
subtract out the large $t-t'$ limit of $\tilde C_a(t-t')$ and use
fixed $x_i^{ab}$ (constant in time and across trials) to generate
the input noise. (If we did it the other way, without the
subtraction, we would effectively be assuming that $x_i^{ab}$
changed randomly from trial to trial, which is not correct.)

In our calculations we have used 10000 trials to calculate these
single-neuron firing statistics.  We perform the subtraction of
the long-time limit of $\tilde C_a(t-t')$ at $|t-t'|= 50$, and we
have checked that (\ref{eq:Ctilde}) is flat beyond this point in
all the cases we have done.

If we perform this kind of measurement separately for many values
of the $x_i^{ab}$, we will be able to see how the firing
statistics vary across the population.  Here, however, we will
confine most of our attention to what we call the ``average
neuron": the one with the average value (0) of all three
$x_i^{ab}$.

In particular, we calculate the mean spike count in the 100-ms
trials and its variance across trials.  From this we can get the
Fano factor $F$ (the variance/mean ratio).  We also compute the
autocorrelation function, which offers a consistency check, since
the Fano factor can also be obtained from \be F =
\frac{1}{r}\int_{-\infty}^{\infty}C(\tau) \d \tau . \lb{eq:FfromC}
\ee (This formula is valid when the measurement period is much
larger than the time over which $C(\tau)$ falls to zero.)

We will study below how these firing statistics vary as we change
various parameters of the model: the input rates $r_0$, parameters
that control the balance of excitation and inhibition, and the
overall strength of the synapses.  This will give us some generic
understanding of what controls the degree of irregularity of the
neuronal firing.

\subsection*{Nonstationary Case}

When the input population is not firing at a constant rate, almost
the same calculational procedure can be followed, except that one
does not average measured rates, their fluctuations or correlation
function over time.  To start out, we get initial instantaneous
rate estimates from the balance condition, assuming that the
time-dependent average input currents do not vary too quickly.
(This condition is not very stringent; van Vreeswijk and
Sompolinsky showed that the stability eigenvalues are proportional
to $\sqrt{K_0}$, so if they have the right sign the convergence to
the balanced state is very rapid.)

To do the iterative procedure to satisfy the self-consistency
conditions of the theory, it is simplest to use the second of the
two ways described above (not doing any subtraction until the
final calculations with single neurons).  In this case the
expression (\ref{eq:Bb}) does not include rate fluctuations, and
we get equations for the noise input currents just like
(\ref{eq:invar}), (\ref{eq:innoise}) and (\ref{eq:decrec}) except
that the $r_b$ are $t$-dependent and the correlation functions
$C_b$ and $D_b$ depend on both $t$ and $t'$, not just their
difference.

The only tricky part is the subtraction of the long-time limit of
the correlation function, which is not simply defined.

We treat this problem in the following way.  We examine the
rate-normalized  quantity \be \hat D_a(t,t') =
\frac{D_a(t,t')}{r_a(t)r_a(t')}.         \lb{eq:Dhat} \ee We find
that this quantity is time-translation invariant (i.e., a function
only of $t-t'$) to a very good approximation, so we perform the
subtraction of the long-time limit on it.  Then multiplying the
subtracted $\hat D$ by $r_a(t)r_a(t')$ gives a good approximation
to the true correlation function $C_a(t,t')$.  The meaning of this
finding is, loosely speaking, that when the rates vary (slowly
enough) in time, the correlation functions just inherit these
rates as overall factors without changing anything else about the
problem.

We will use the this time-dependent formulation below to simulate
experiments like those of Gershon {\em et al} \cite{Gershonetal},
where the LGN input $r_0(t)$ to visual cortical cells is
time-dependent because of the flashing-on and off of the stimulus.

\section{Results}

The results presented in this chapter were obtained from
simulations with parameters corresponding to population sizes of
$N_1 =$ 40,000 excitatory neurons and $N_2 =$ 10,000 inhibitory
neurons. With the above mentioned connection probabilities of
$K_a/N_a = 0.1$, this translates to an average number of $K_1 =
4000$ excitatory inputs and $K_2 = 1000$ inhibitory inputs to each
neuron. The average number of external (excitatory) inputs $K_0$
was chosen to be equal to $K_2$. All neurons have the same
membrane time constant $\tau$ of $10$~ms.

To study the effect of various combinations in synaptic strength,
we use the following generic form to define the intra-cortical
weights $J_{ab}$:

\be
    \left( \begin{array}{cc}
        J_{11} & J_{12} \\
        J_{21} & J_{22}
    \end{array} \right) =
    \left( \begin{array}{cc}
        \epsilon & -2g \\
        1        & -2g
    \end{array} \right)             \lb{eq:Jmatrix}
\ee For the synaptic strengths from the external population we use
$J_{10} = 1$ and $J_{20} = \epsilon$. With this notation, $g$
determines the strength of inhibition relative to excitation
within the network, and $\epsilon$ the strength of intracortical
excitation. Additionally, we scale the overall strength of the
synapses with a multiplicative scaling factor denoted $J_s$ so
that each synapse has an actual weight of $J_s \cdot J_{ab}$,
regardless of $a$ and $b$.

\begin{figure}[t]
\centering
\begin{tabular}{c}
\begin{minipage}{10cm}
   \includegraphics[width=9cm,height=5.4cm]{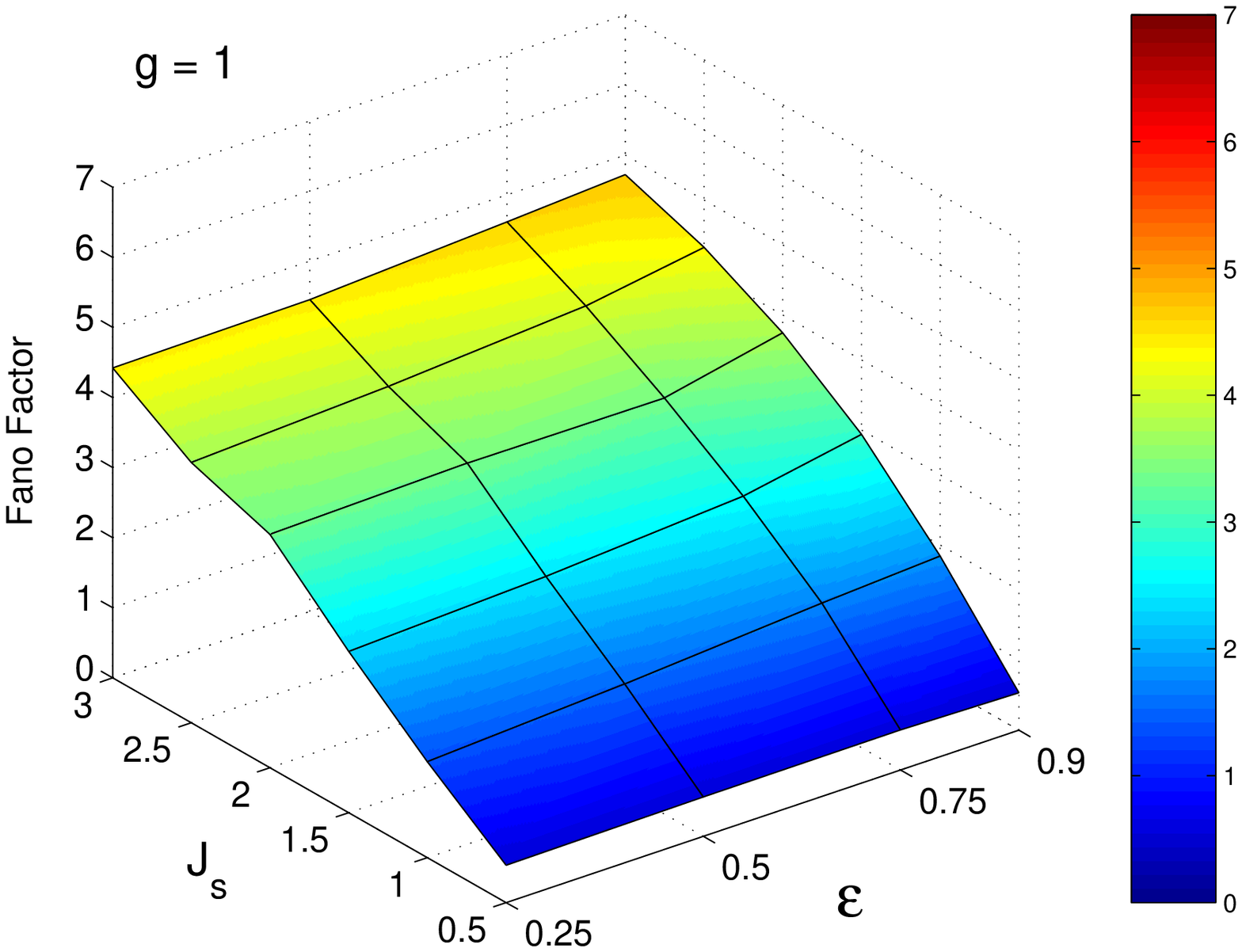}
\end{minipage}
\\
\begin{minipage}{10cm}
   \includegraphics[width=9cm,height=5.4cm]{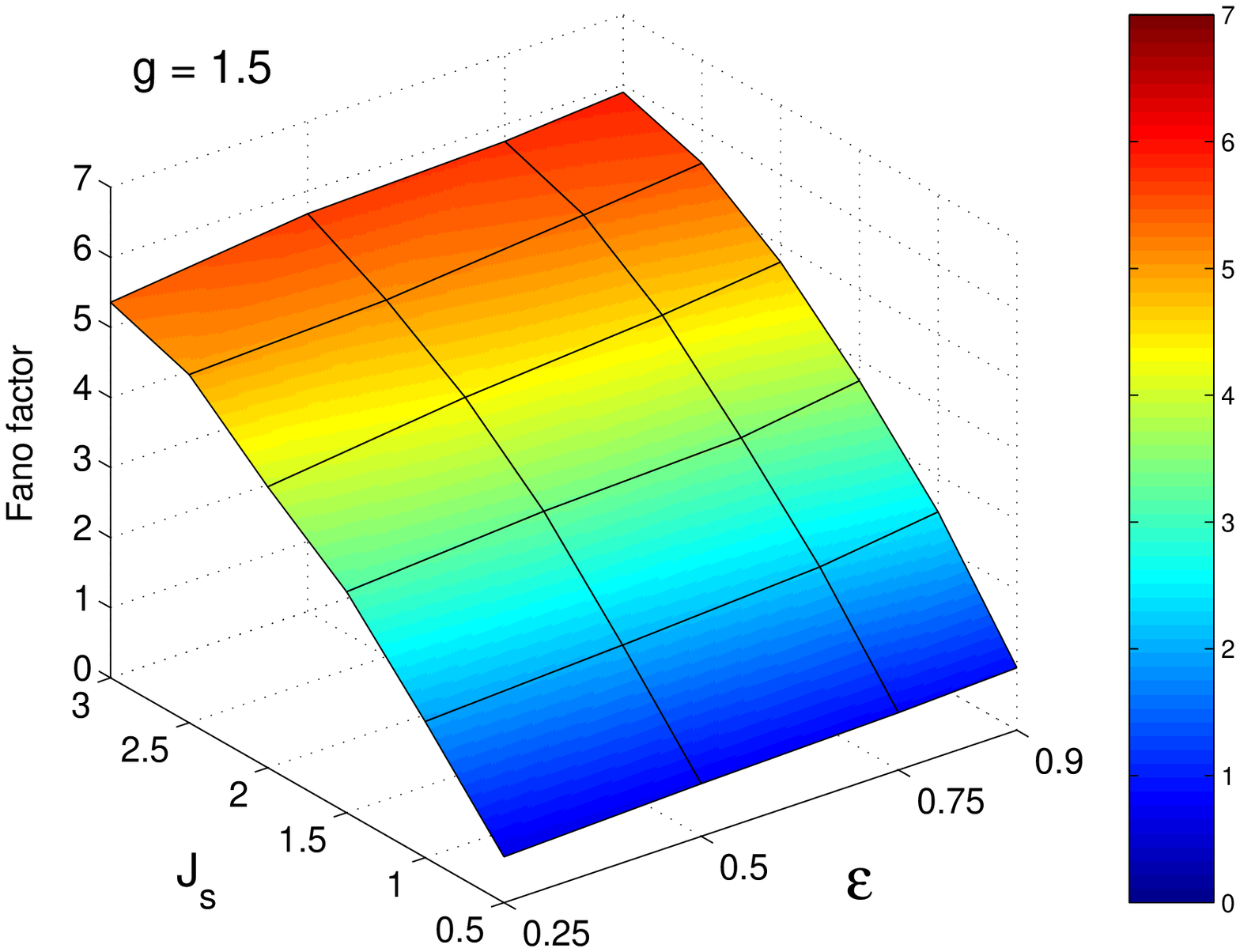}
\end{minipage}
\\
\begin{minipage}{10cm}
   \includegraphics[width=9cm,height=5.4cm]{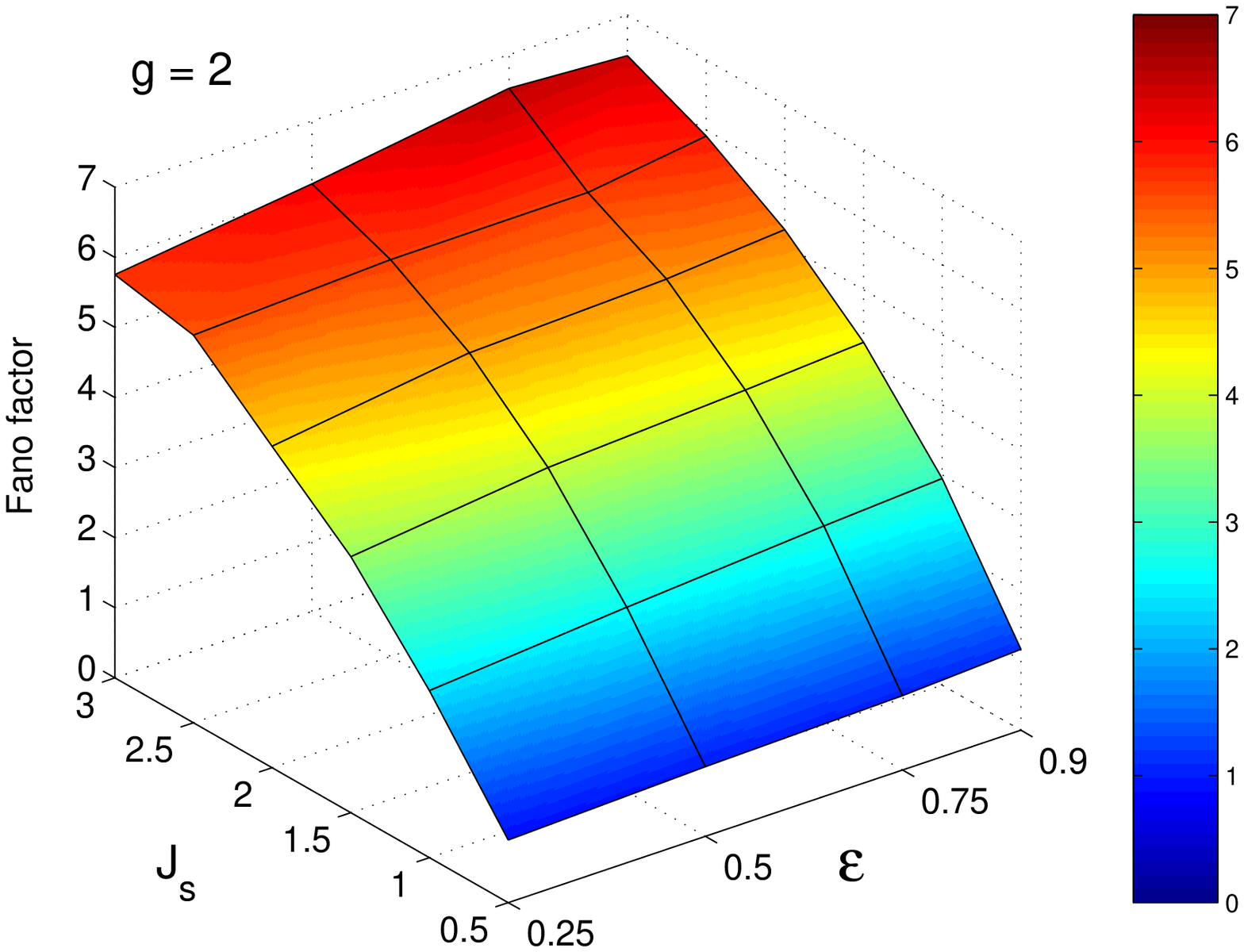}
   \caption{Fano factors as a function of overall synaptic strength $J_s$ and
   intracortical excitation strength $\epsilon$ for three different inhibition
   factors: $g = 1$, $1.5$, and $2$, respectively. The increase of any of
   these parameters results in more irregular firing statistics as measured by
   the Fano factor.}
   \label{fig:Fanos}
\end{minipage}
\end{tabular}
\end{figure}

Figure~\ref{fig:Fanos} summarizes how the firing firing statistics
depend on all of the parameters $g$, $\epsilon$, and $J_s$. The
irregularity of spiking, as measured by the Fano factor, depends
most sensitively on the overall scaling of the synaptic strength,
$J_s$. The Fano factor increases systematically as $J_s$
increases, and higher values of intracortical excitation
$\epsilon$ also result in higher values of $F$. The same pattern
holds for stronger intracortical inhibition, parameterized by $g$.
For all of these cases the mean firing rate remains virtually
unchanged due to the dynamic balance of excitation and inhibition
in the network, whereas the fluctuations increase with the
increase of any of the synaptic weights.

\begin{figure}[t]
\centering
\begin{tabular}{c}
\begin{minipage}{9cm}
   \includegraphics[width=8.2cm,height=5.4cm]{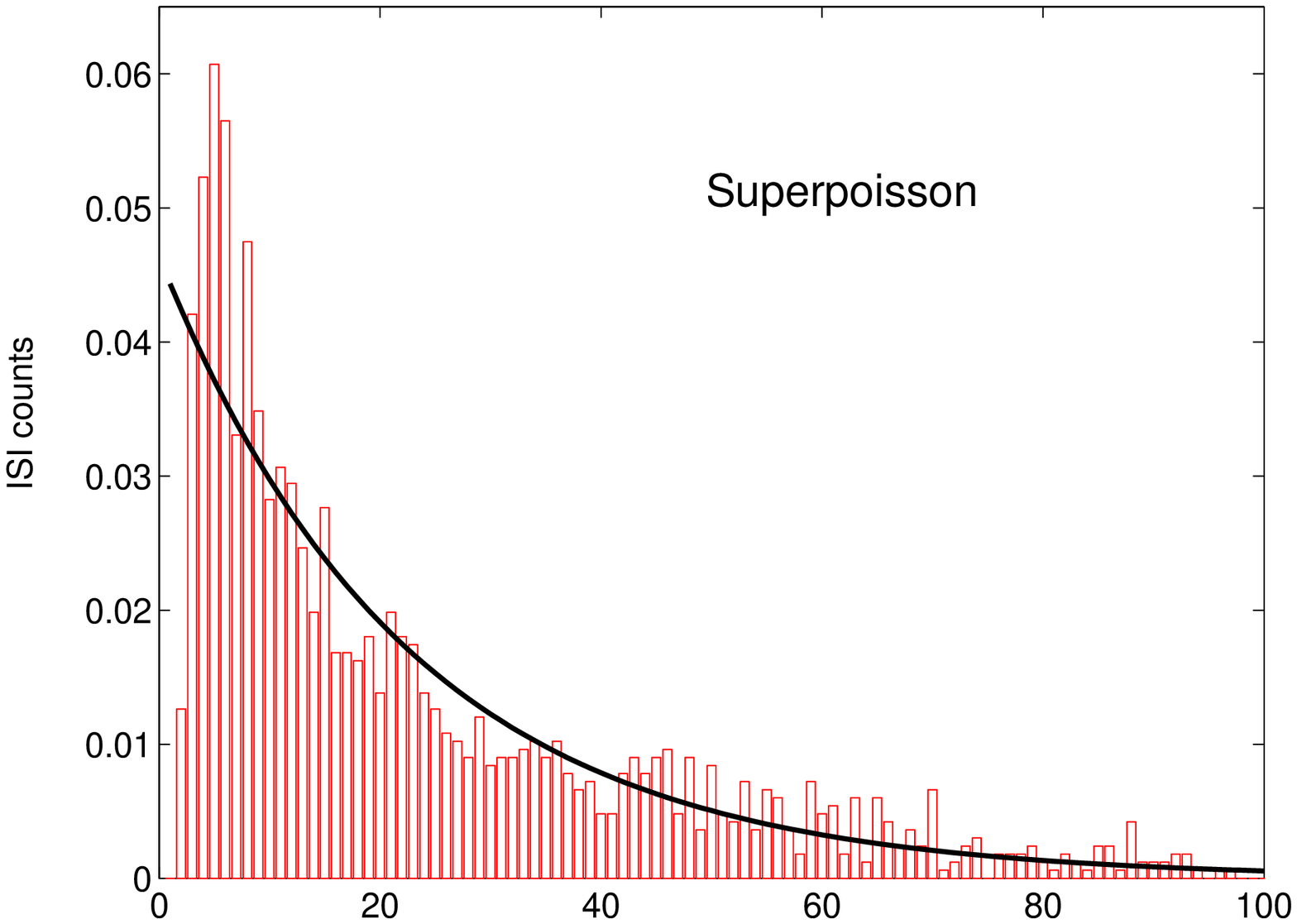}
\end{minipage}
\\
\begin{minipage}{9cm}
   \includegraphics[width=8.2cm,height=5.4cm]{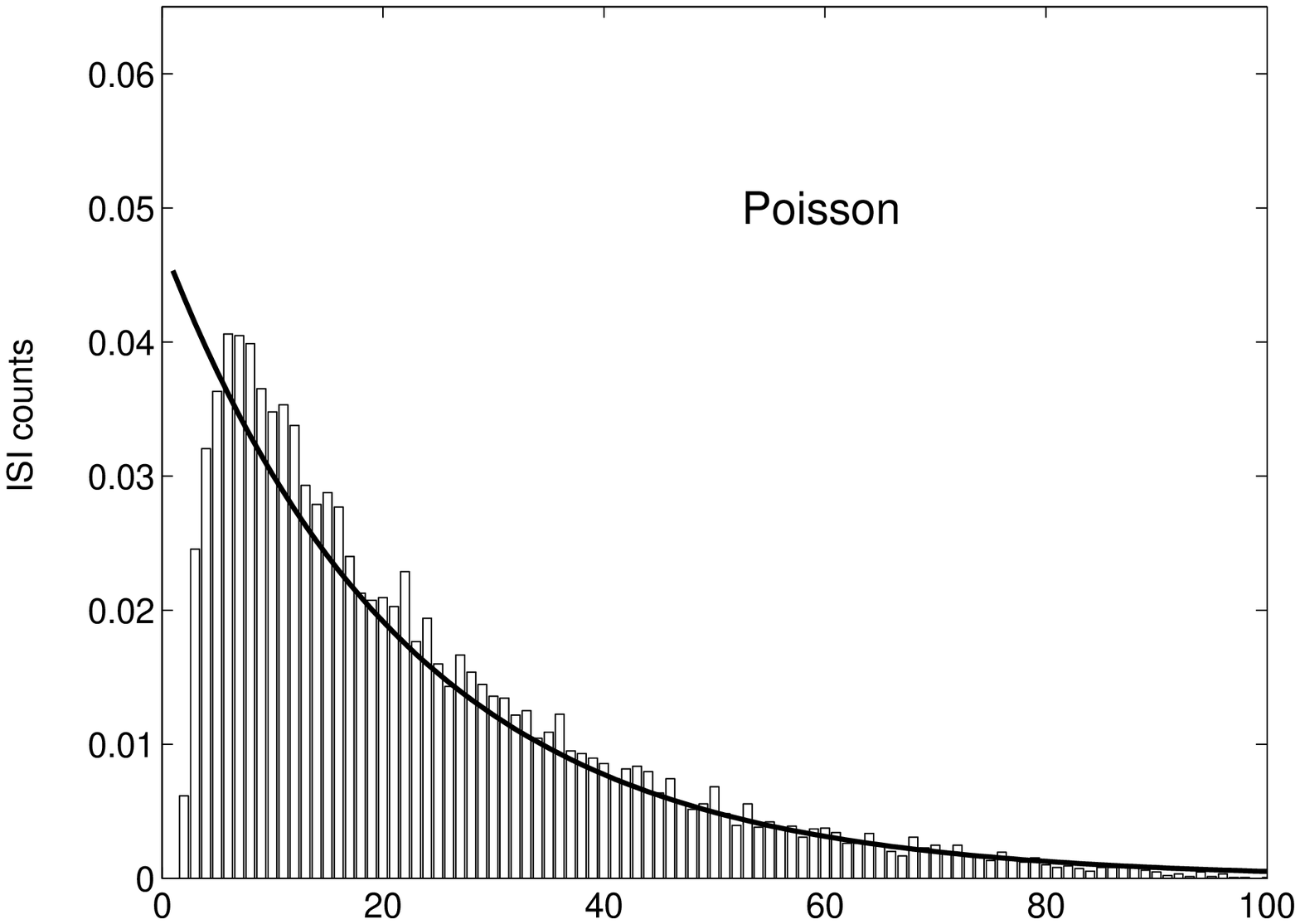}
\end{minipage}
\\
\begin{minipage}{9cm}
   \includegraphics[width=8.2cm,height=5.4cm]{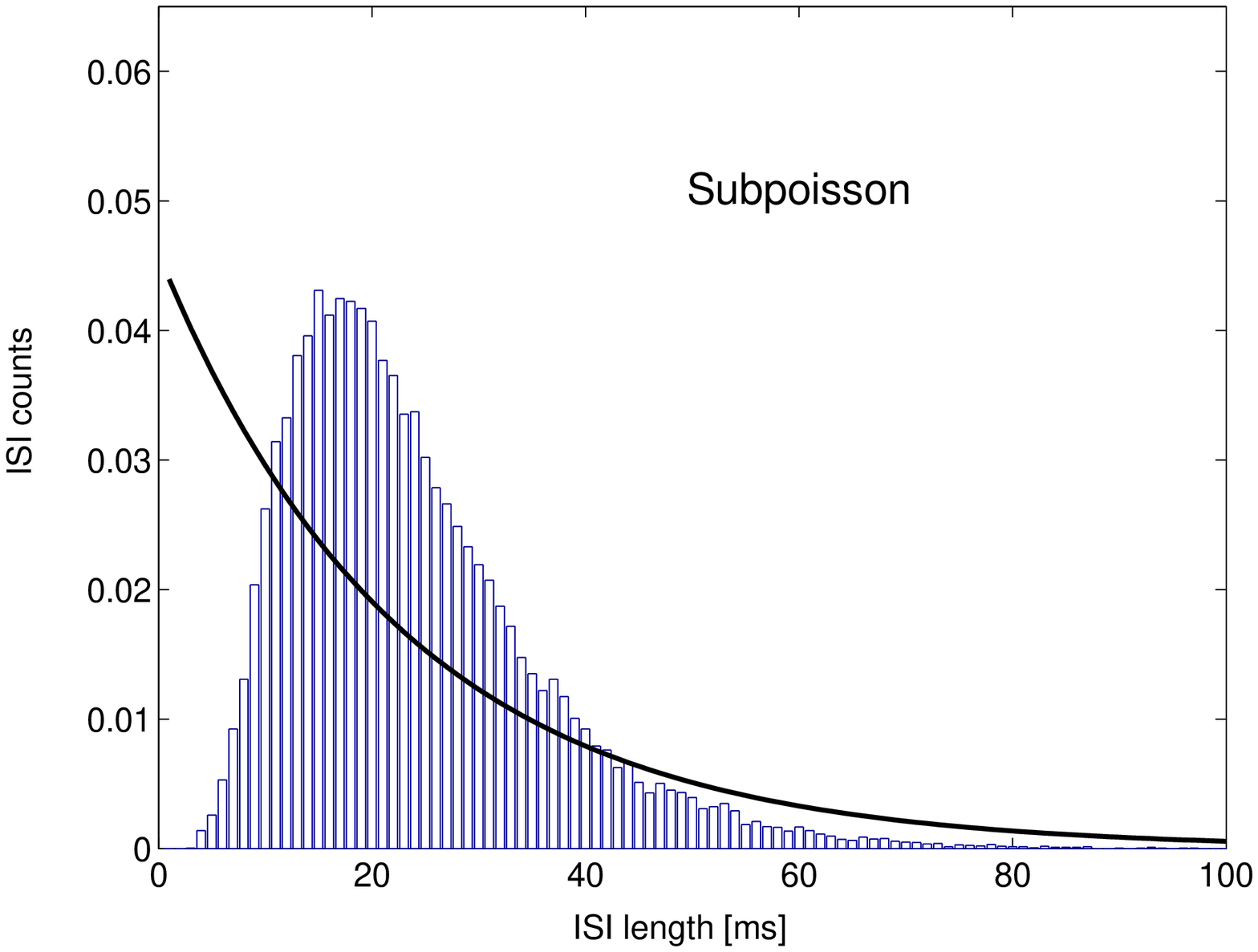}
   \caption{Interspike interval distributions for fixed $\epsilon = 0.5$ and
   $g = 1$, and three different values of overall synaptic strength $J_s$:
   $1.5$ (superpoissonian), $0.75$ (Poissonian), and $0.375$ (subpoissonian).
   Overlayed on each figure is the exponential fall-off of a true Poisson
   distribution with the same average rate as in all of the three cases.}
   \label{fig:ISI}
\end{minipage}
\end{tabular}
\end{figure}

Interspike interval (ISI) distributions are shown in
Figure~\ref{fig:ISI} for three different values of $J_s$, keeping
$\epsilon$ and $g$ fixed at $0.5$ and $1$, respectively. For a
Poisson spike train, the Fano factor $F = 1$, while $F > 1$ (which
we term ``superpoissonian") indicates a tendency of spikes
occurring in clusters separated by accordingly longer empty
intervals, and $F < 1$ (``subpoissonian") indicates more
regularity, reflected by a narrower distribution.  We have
adjusted the input rate $r_0$ so that the output rate is the same
in all three cases.

The top panel of Figure~\ref{fig:ISI} shows the ISI distribution
of a superpoissonian spike train, obtained for $J_s = 1.5$.
Overlayed on the histogram of ISI counts is an exponential curve
indicating a Poisson distribution with the same mean ISI length.
Compared with the Poisson distribution, the superpoissonian spike
train contains more short intervals, as seen by the peak at short
lengths, and also more long intervals, causing a long tail.
Necessarily, the interval count around the average ISI length is
lower than that for a Poisson spike train.

The ISI distribution in the middle panel of Figure~\ref{fig:ISI}
belongs to a spike train with a Fano factor close to one, obtained
for $J_s = 0.75$. The overlayed exponential reveals a deviation
from the ISI count: while intervals of diminishing length are the
most likely ones for a real Poisson process, our neuronal spike
trains always show some refractoriness reflected by a dip at the
shortest intervals.  (We have not used an explicit refractory
period in our model.  The dip seen here simply reflects the fact
that it takes a little time for the membrane potential
distribution to return to its steady-state form after reset.)
Apart from this deviation, however, there is a close resemblance
between the observed distribution and the ``predicted" one.

Finally, the lower panel of Figure~\ref{fig:ISI} depicts a case
with $F < 1$, with weaker synapses, leading to a stronger
refractory effect and (since the rate is fixed) an accordingly
narrower distribution around the average ISI length, as compared
to the overlayed Poisson distribution. This distribution was
obtained with weak synapses produced by a small scaling factor of
$J_s = 0.375$.

\begin{figure}[t]
\centering
 \includegraphics[width=10.2cm,height=7cm]{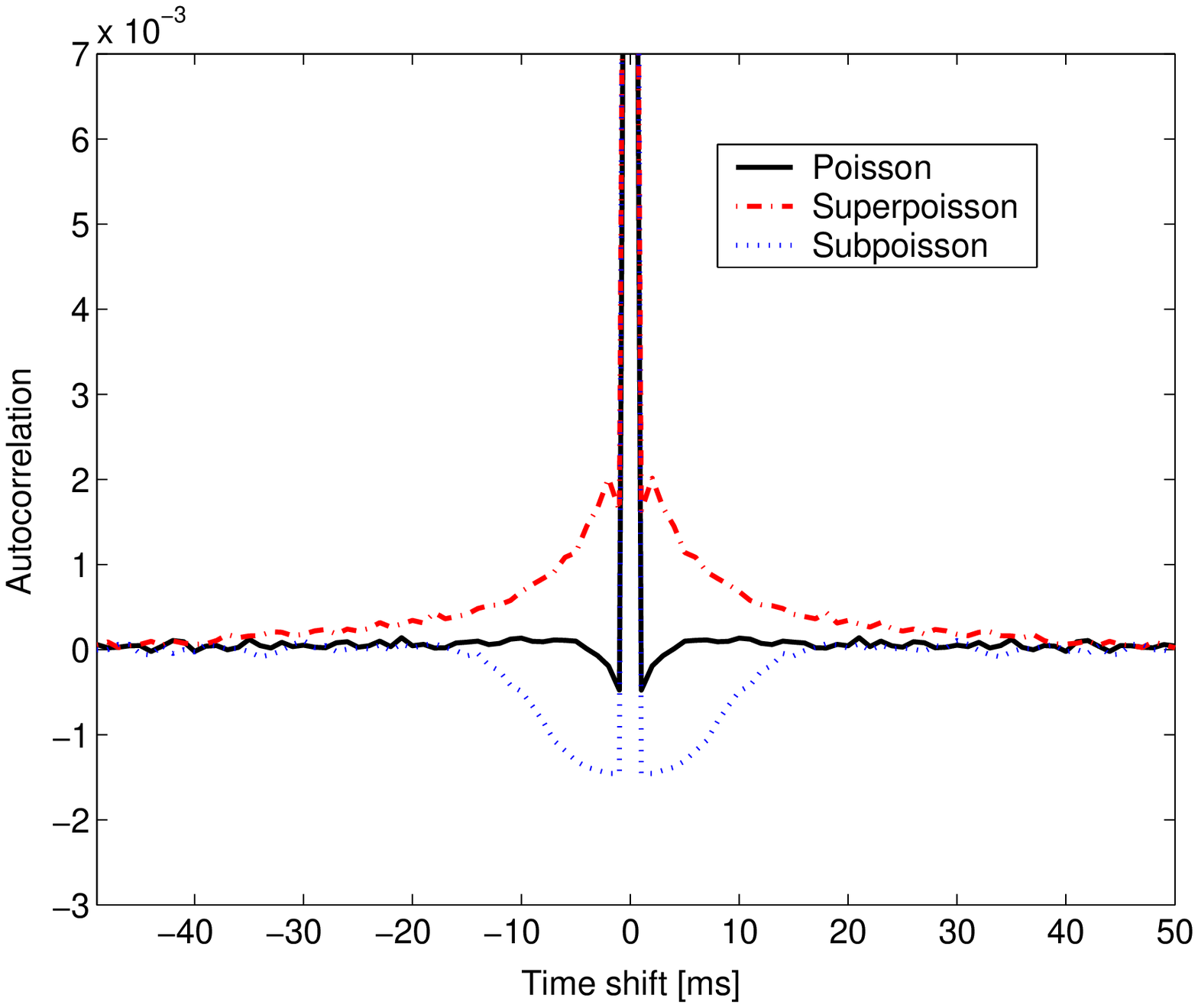}
\caption{Three different spike train autocorrelations illustrating
the relationship between the Fano factor $F$ and the area under
the curve. For $F = 1$ (Poissonian, black solid line), the
autocorrelation is an almost perfect delta function. $F > 1$
(superpoissonian, red dash-dot line) is reflected by a hill
generating a positive area, and $F < 1$ (subpoissonian, blue
dotted line) is accompanied by a valley of negative correlations.
(See the text for more details.)} \label{fig:AutoCorr}
\end{figure}

As mentioned in the previous section, the Fano factor can also be
obtained by integrating over the spike train autocorrelation
divided by the spike rate (\ref{eq:FfromC}). For a Poisson process
the autocorrelation vanishes for all lags different from zero. In
contrast, $F>1$ (superpoissonian case) implies a positive integral
over non-zero lags, whereas in the subpoissonian case there must
be a negative area under the curve. Figure~\ref{fig:AutoCorr}
shows examples of autocorrelations for all of the three cases. For
the superpoissonian case (red dash-dot line), there is a ``hill"
of positive correlations for short intervals, reflecting the
tendency toward spike clustering. The subpoissonian
autocorrelation (blue dotted line) shows a valley of negative
correlations for short intervals, indicating well separated spikes
in a more regular spike train. The curve labeled as Poisson (black
solid line) does have a small valley around zero lag, which
reflects once more the refractoriness of neurons to fire at
extremely short intervals, unlike a completely random Poisson
process.  (Actually, the measured $F$ in this case is slightly
greater than 1, implying that in this case the integral of the
very small positive tail for $t > 2$ ms is slightly larger than
that of the (more obvious) negative short-time dip.)


\begin{figure}[t]
\centering
 \includegraphics[width=9cm,height=7cm]{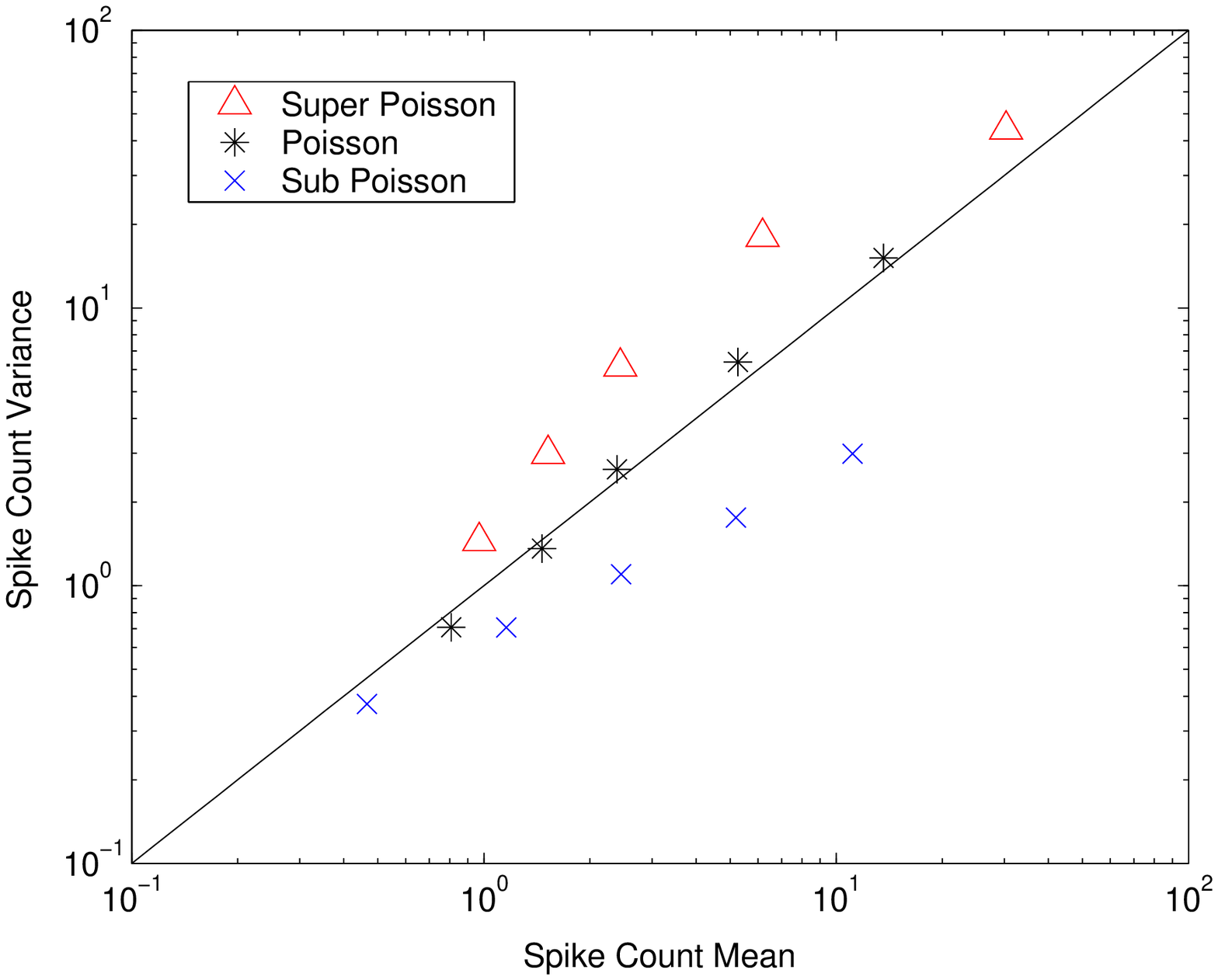}
\caption{Spike count log(variance) vs.\ log(mean) for three
different values of overall synaptic strength $J_s$, varying the
external input rate $r_0$. For $J_s = 1.25$ (superpoissonian, red
triangles) the data look qualitatively like those from
experiments. The other values for $J_s$ are $0.75$ (Poisson, black
stars) and $0.375$ (subpoissonian, blue crosses).}
\label{fig:VarMean}
\end{figure}

Measurements on V1 neurons in awake monkeys (see for example
Gershon \emph{et al.}~\cite{Gershonetal}) suggest a linear
relationship between the log variance and the log mean of
stimulus-elicited spike counts. We find a similar dependence for
neurons within our model network. Figure~\ref{fig:VarMean} shows
results for three different values of $J_s$.  In each  case, five
different values of the external input rate $r_0$ were tried,
causing various mean spike counts and variances. The logarithm of
the spike count variance is plotted as a function of the logarithm
of the spike count mean, and a solid diagonal line indicates the
identity, i.e, a Fano factor of exactly 1.  We see that for the
largest value of $J_s$ used here, the data look qualitatively like
those from experiments, with Fano factors in the range around 1.5
to 2.

\subsection*{Nonstationary Case}

\begin{figure}[t]
\centering
 \includegraphics[width=9cm]{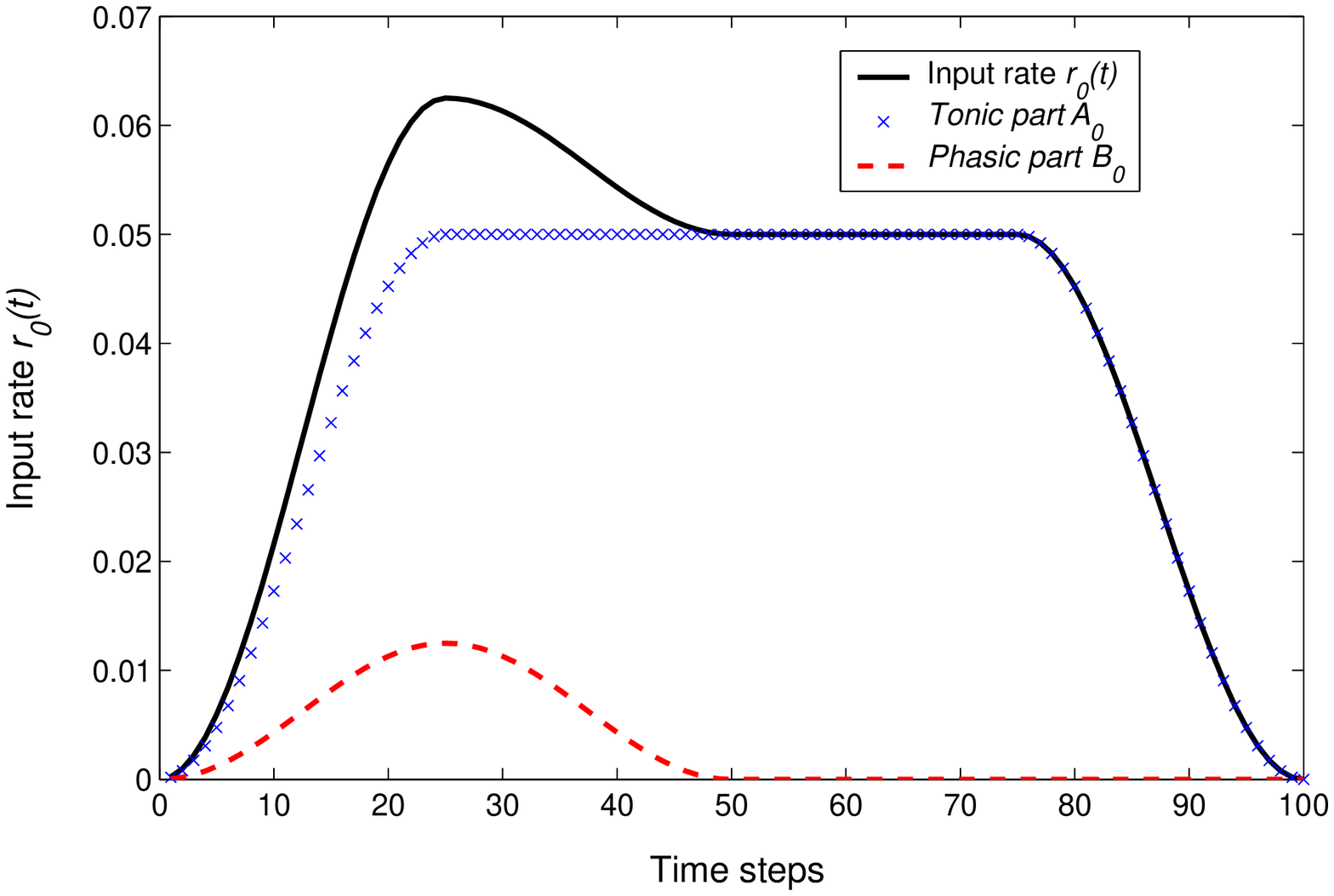}
\caption{Parametrization of the time-dependent input rate
$r_0(t)$. The input is modeled as the sum of three functions: (1)
a stationary background rate (which is zero in this case); (2) a
tonic part, which rises within the first 20 ms to a constant level
of $A_0$ where it stays for 60 ms, falling back to zero within the
last 20 ms; and (3) an initial phasic part, which is nonzero only
in the first 50 ms, rising to a maximum value of $B_0$.}
\label{fig:NonstatInput}
\end{figure}

The results presented in the previous section were obtained with
stationary inputs, while experimental data like those from
\cite{Gershonetal} were collected from visual neurons subject to
time-dependent inputs. Therefore, we performed calculations of the
spike statistics in which the input population rate $r_0$ was
time-dependent. The modeled temporal shape of $r_0(t)$ is depicted
in Figure~\ref{fig:NonstatInput}. It is the sum of three terms:
\begin{equation}\label{eq:r0(t)}
    r_0(t) = R_0 + A(t) + B(t)
\end{equation}
The first, $R_0$, is just a constant, as in the preceding section.
The second term, $A(t)$, rises to a maximum over a 25-ms interval,
remains constant for 50~ms, and then falls off to zero over the
final 25~ms.
\begin{equation}\label{eq:A(t)}
    A(t) = \left\{%
\begin{array}{ll}
    0.5 A_0 (1-\cos(4t\pi/T))      & \hbox{for $0 < t \leq T/4$} \\
    A_0                            & \hbox{for $T/4 < t \leq 3T/4$} \\
    0.5 A_0 (1-\cos(4(T-t)\pi/T))  & \hbox{for $3T/4 < t \leq T$,} \\
\end{array}%
\right.
\end{equation}
where $T$ is the total simulation interval of 100~ms. The third
term, $B_0$, rises to a maximum in the first 25~ms and then falls
back to zero in the next 25~ms, remaining zero thereafter.
\begin{equation}\label{eq:B(t)}
    B(t) = \left\{%
\begin{array}{ll}
    0.5 B_0 (1-\cos(4t\pi/T))        & \hbox{for $0 < t \leq T/4$} \\
    0.5 B_0 (1-\cos(4(T/2-t)\pi/T))  & \hbox{for $T/4 < t \leq T/2$} \\
    0                                & \hbox{for $T/2 < t \leq T$.} \\
\end{array}%
\right.
\end{equation}

\begin{figure}[t]
\centering
 \includegraphics[width=9cm,height=7cm]{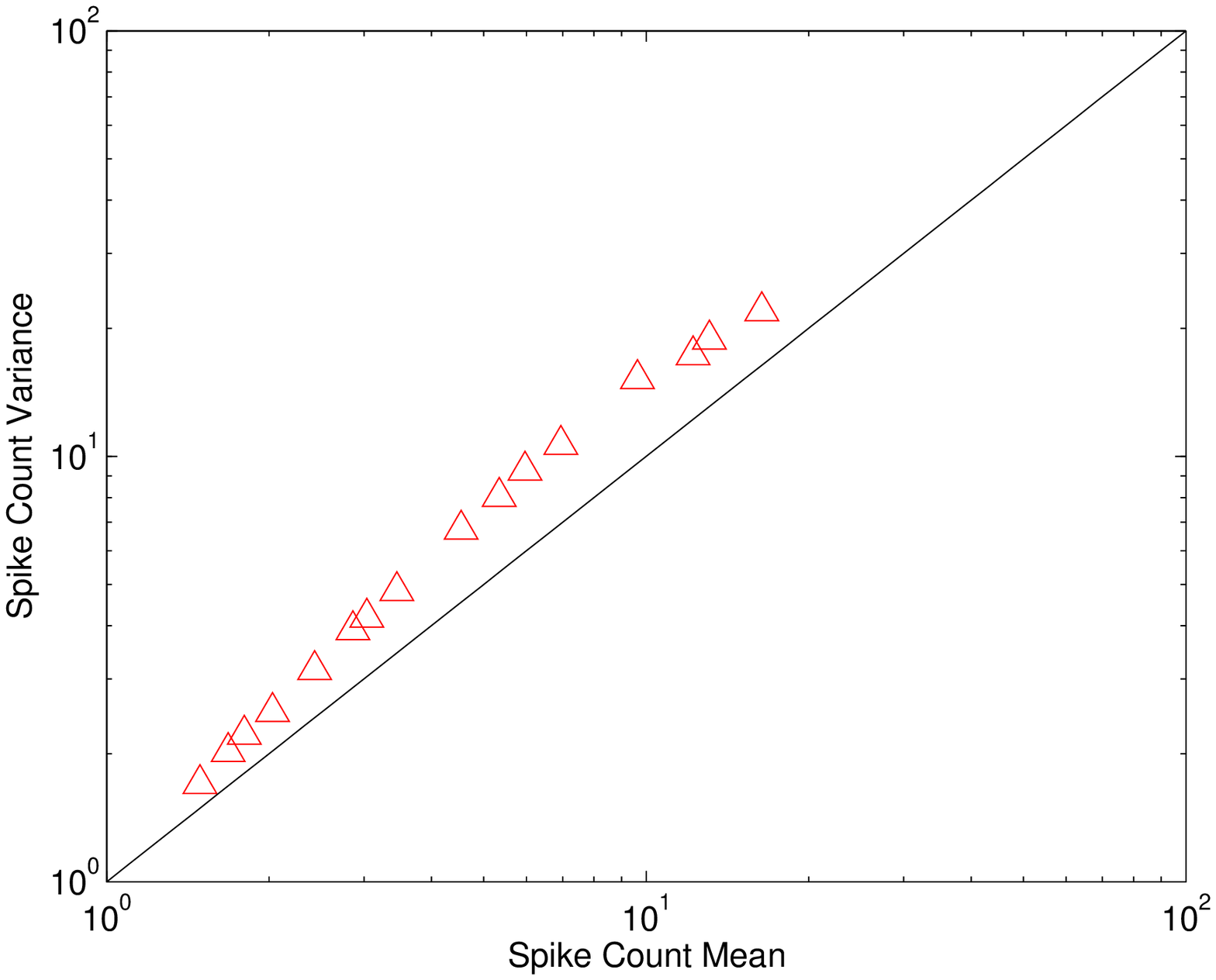}
\caption{Spike count log(variance) vs.\ log(mean) for time-varying
external inputs with varying overall strength. The neuron in the
simulated network (red triangles) fires in a superpoissonian
regime, with an almost linear relationship for low spike rates
between the log variance and the log mean, resembling closely data
obtained from in-vivo experiments. The diagonal solid line
indicates the identity of variance and mean (Fano factor $F =
1$).} \label{fig:NonstatVarMean}
\end{figure}

Figure~\ref{fig:NonstatVarMean} shows the logarithm of the spike
count variance plotted against the logarithm of the spike count
mean for various non-stationary inputs characterized by different
values of $A_0$ and $B_0$. The graph shows results for $J_s = 1$,
$\epsilon = 0.5$, $g = 1$, and a background rate of $R_0 = 0.1$.
Table~\ref{tab:nonstat-results} shows the choice of the sixteen
combinations of the stimulus parameters $A_0$ and $B_0$, together
with the resulting Fano factors $F$ for the simulated neuron.

\begin{table}
  \centering
\begin{tabular}{|c|c|c|c|c|c|c|c|c|}
  \hline
  $A_0$ & $0.375$ & $0.375$ & $0.500$ & $0.500$ & $0.750$ & $0.750$ & $1.000$ & $1.000$ \\
  $B_0$ & $0.125$ & $0.375$ & $0.125$ & $0.375$ & $0.250$ & $0.750$ & $0.250$ & $0.750$ \\
  \hline
  $F$ & $1.14$ & $1.2$ & $1.22$ & $1.23$ & $1.29$ & $1.36$ & $1.37$ & $1.4$  \\
  \hline
\end{tabular}
\begin{tabular}{|c|c|c|c|c|c|c|c|c|}
  \hline
  $A_0$ & $1.500$ & $1.500$ & $2.000$ & $2.000$ & $3.000$ & $3.000$ & $4.000$ & $4.000$\\
  $B_0$ & $0.500$ & $1.500$ & $0.500$ & $1.500$ & $1.000$ & $3.000$ & $1.000$ & $3.000$ \\
  \hline
  $F$ & $1.48$ & $1.5$ & $1.55$ & $1.53$ & $1.57$ & $1.41$ & $1.43$ & $1.34$\\
  \hline
\end{tabular}
\caption{Stimulus parameters $A_0$ and $B_0$ for the results
depicted in Figure~\ref{fig:NonstatVarMean}, and the resulting
Fano factors $F$.}\label{tab:nonstat-results}
\end{table}

The data look qualitatively like those obtained from
\emph{in-vivo} experiments \cite{Gershonetal} and are similar to
the superpoissonian case in Figure~\ref{fig:VarMean}. The neuron
fires consistently in a superpoissonian regime with Fano factors
slightly higher than 1 and an almost linear relationship between
the log variance and the log mean for low spike counts. For higher
spike counts, the curve bends towards values of lower Fano
factors, just as for stationary inputs (Figure~\ref{fig:VarMean}).
In both cases, this bend reflects the the decrease in irregularity
of firing caused by an increasingly prominent role of
refractoriness for shorter interspike intervals.

\section{Discussion}

Cortical neurons receive thousands of both excitatory and
inhibitory inputs, and despite the high number of inputs from
nearby neurons with similar firing statistics and similar
connectivity, their observed firing is very irregular
\cite{HeggelundAlbus,Dean,TMT,TMD,Vogelsetal,Snowdenetal,Guretal,
ShadlenNewsome,Gershonetal,Karaetal, Buracasetal,
Leeetal,DeWeeseetal}. Dynamically balanced excitation and
inhibition through a simple feedback mechanism provides an
explanation that naturally accounts for this phenomenon without
requiring fine tuning of the parameters
\cite{AmitBrunelCC,AmitBrunelNetwork,BrunelJCNS, vVSScience,
vVSNC}. Moreover, neurons in such model networks show an almost
linear input-output relationship (input current versus firing
frequency), as do neurons in the neocortex.

Here, we have extended the mean-field description of the
dynamically balanced asynchronous firing state to analyze firing
correlations. We found that the relationship between the observed
irregularity of firing (spike count variance) and the firing rate
(spike count mean) of the neurons resemble closely data collected
from \emph{in-vivo} experiments (see Figures \ref{fig:VarMean} and
\ref{fig:NonstatInput}). To do this, we developed a complete
mean-field theory for a network of leaky integrate-and-fire
neurons, in which both firing rates and correlation functions are
determined self-consistently. Using an algorithm that allows us to
find the solutions to the mean-field equations numerically, we
could elucidate how the strength of synapses within the network
influences the expected firing statistics of cortical neurons in a
systematic manner (see Figure~\ref{fig:Fanos}).

We have shown that the irregularity of firing, as measured by the
Fano factor, increases with increasing synaptic strengths
(Figure~\ref{fig:Fanos}). Nearly Poisson statistics (with $F
\approx 1$) are observed for moderately strong strengths, but the
transition from subpoissonian to superpoissonian statistics is
smooth, without a special role for $F = 1$.

The higher irregularity in the spike counts is always accompanied
by a tendency toward more ``bursty" firing. (These bursts are a
network effect; the model contains only leaky integrate-and-fire
neurons, which do not burst on their own.) This burstiness can
best be seen in the spike train autocorrelation function
(Figure~\ref{fig:AutoCorr}), which acquires a hill of growing size
and width around zero lag for increasing Fano factors. The
interdependence between firing irregularity and bursting can be
understood with help of the ISI distributions depicted in
Figure~\ref{fig:ISI}: when the rate, and thus the average ISI, is
kept constant, then any higher count for shorter-than-average ISIs
must be accompanied by an accordingly higher count for longer ISIs
(indicating bursts), and vice versa. Thus higher irregularity
always goes hand in hand with a higher tendency toward temporal
clustering of spikes.

Why do stronger synapses lead to higher irregularity in firing?
The size of the input current fluctuations in (\ref{eq:decrec})
are controlled by the $J_{ab}$, and so, therefore, are the
corresponding membrane potential fluctuations.  Thus, for example,
the width of the steady-state membrane potential distribution is
proportional to $J_s$.  We next have to consider where this
distribution is centered.  Remembering that, according to the
balance condition, the firing rate is independent of $J_s$, the
center of the distribution has to move farther away from threshold
as $J_s$ is increased in order to keep the rate fixed.  Therefore,
for very small $J_s$ almost the entire equilibrium membrane
potential distribution will lie well above the post-spike reset
value, while for large $J_s$ it will be mostly below reset.

Immediately after a spike, the membrane potential distribution is
a delta-function centered at the reset (here 0).  It then spreads
and its mean moves up or down toward its equilibrium value.  This
equilibration will take about a membrane time constant. If the
equilibrium value is well above zero (the small-$J_s$ case), the
probability of reaching threshold will be suppressed during this
time, implying a refractory dip in the ISI distribution and the
correlation function and a tendency toward a Fano factor less than
1.

In the large-$J_s$ case, on the other hand, where the membrane
potential is reset much closer to the threshold than to its
eventual equilibrium value, the initial rapid spread (with the
width growing proportional to $J_s \sqrt{t}$) leads to an enhanced
probability of early spikes.  At short times this diffusive spread
dominates the downward drift of the mean (which is only linear in
$t$).  Thus there is extra weight in the ISI distribution and a
positive correlation function at these short times, leading to a
Fano factor greater than 1.

Empirically, an approximate power-law relationship between the
mean and variance of the spike count has frequently been observed
for cortical neurons (see, e.g., \cite{TMT, Vogelsetal,
Gershonetal, Leeetal}). Our model shows the same qualitative
feature (Figures \ref{fig:VarMean} and \ref{fig:NonstatInput}),
though we have no argument that the relation should be an exact
power law. However, this agreement suggests that the model
captures at least part of physics underlying the firing
statistics.

As already observed, not all of the variability in measured neuron
responses has to be explained in the manner outlined above.
Changing conditions during the run of a single experiment may
introduce extra irregularity, caused by collecting statistics over
trials with different mean firing rates. The present analysis
shows why -- and how much -- irregularity can be expected due to
intrinsic cortical dynamics.

Our formulation of the mean-field theory is general enough to
allow straightforward extensions to greater biological realism and
to more complicated network architectures. We have introduced a
generalization of this model with conductance-based synapses in
another paper \cite{Lerchner+Ahmadi+Hertz:2004}. We have also
extended the model to include systematic structure in the
connections, modeling an orientation hypercolumn in the primary
visual cortex \cite{Hertz+Sterner:2003}. Moreover, our algorithm
for finding the mean-field solutions is not restricted to networks
of integrate-and-fire neurons. It can be applied to any kind of
neuronal model. Furthermore, any kind of synaptic dynamics can be
incorporated by using synaptically filtered spike trains to
compute the self-consistent solutions.

\end{document}